\documentclass[a4paper,12pt]{article}
\usepackage{multicol}
\usepackage[left=1.5cm,right=1.5cm,top=2cm,bottom=2cm]{geometry}

\usepackage{cmap}					
\usepackage{ccaption} 
\captiondelim{. } 
\usepackage{cite}

\usepackage{amsfonts,amssymb,amsthm,mathtools} 
\usepackage{amsmath}
\usepackage{icomma} 

\usepackage{euscript}	 
\usepackage{mathrsfs} 
\usepackage{txfonts}

\usepackage{graphicx}  
\usepackage{subfig}
\usepackage{float}
\graphicspath{{images/}}  
\usepackage{wrapfig} 

\usepackage{array,tabularx,tabulary,booktabs} 
\usepackage{longtable}  
\usepackage{multirow} 

\title{EMISSION OF HIGH-ENERGY GAMMA QUANTA BY ULTRARELATIVISTIC ELECTRONS ON NUCLEI IN STRONG X-RAY FIELDS}
\author{S. P. Roshchupkin$^1$, A.V. Dubov$^{1,2}$,  S. S. Starodub$^3$\\\\
$^1$ Peter the Great St. Petersburg Polytechnic University,\\
195251, St-Petersburg, Russian Federation, Russia\\
$^2$ Aalto University, FI-00076 AALTO, Helsinki, Finland,\\
$^3$ Institute of Applied Physics, National Academy of Sciences of Ukraine,\\
 40000, Sumy, Ukraine} 
 
\date{}

\begin{document} 

\maketitle


\begin{abstract}
The possibility of radiation of high-energy gamma quanta with energies of the order of 100 GeV by ultrarelativistic electrons on nuclei in strong X-ray fields with intensities up to $\sim 10^{27}\ \text{Wcm}^{-2}$ has been theoretically studied. It is shown that this effect can be realized under special experimental conditions in the process of resonant spontaneous bremsstrahlung radiation of ultrarelativistic electrons on nuclei in an external electromagnetic field. These special experimental conditions determine the characteristic energy of the electrons. This characteristic energy should be significantly less than the energy of the initial electrons. Under these conditions, spontaneous gamma quanta are emitted in a narrow cone with energies close to the energy of the initial electrons. Moreover, the resonant differential cross-sections of such processes can exceed the corresponding differential cross-section without an external field by twenty orders of magnitude. The results obtained can explain the occurrence of high-energy gamma quanta near pulsars and magnetars.
\end{abstract}

\section{Introduction}
Astroparticle physics has been attracting attention for a long time (see, for example, articles \cite{1,2,3,4}). At the same time, the processes of quantum electrodynamics (QED) in X-ray fields near pulsars and magnetars are intensively studied \cite{5,6,7,8,9,10}. We also note a fairly large number of articles on QED processes in strong electromagnetic fields (see, for example, articles \cite{11,12,13,14,15,16,17,18,19,20,21,22,23,24,25,26,27,28,29,30,31,32,33,34,35,36,37,38,39,40,41,42,43}, reviews \cite{44,45,46,47,48,49} and monographs \cite{50,51,52}). It is important to emphasize that higher-order QED processes with respect to the fine structure constant in an electromagnetic field can proceed both in a resonant and non-resonant way. Here, the so-called Oleinik resonances may occur \cite{11,12}, due to the fact that lower-order processes are allowed in the laser field by the fine structure constant. It is important to note that the resonant differential cross-section can significantly exceed the corresponding non-resonant one  \cite{30,31,32,33,47,48,49,50,51,52}.

Resonant spontaneous bremsstrahlung radiation of ultrarelativistic electrons by nuclei in an electromagnetic field was studied in articles \cite{5,31,32,33}. At the same time, unlike the articles \cite{5,30,31,32}, in the article \cite{33} this process was considered in a strong electromagnetic field of optical frequencies. However, these articles did not study the case of strong fields of the X-ray frequency range, when spontaneous gamma quanta are emitted with energies close to the energies of the initial electrons. It is this, the most interesting case, that will be considered in this article.

We will consider resonant processes for high-energy particles when the main parameter is the classical relativistic-invariant parameter
\begin{equation} \label{eq1}
	\eta =\frac{eF\lambdabar}{mc^2}\gtrsim 1,
\end{equation}
numerically equal to the ratio of the field work at the wavelength to the rest energy of the electron ($e$ and $m$ are the charge and mass of the electron, $F$ and\quad$\lambdabar=c/\omega$ are the electric field strength and wavelength, $\omega $ is the frequency of the wave).

In the article \cite{33} it was shown that the resonant frequency of a spontaneous gamma quantum is determined by its outgoing angle relative to the initial electron momentum of the (for channel A) or the final electron momentum (for channel B), as well as the quantum parameter
\begin{equation} \label{eq2}
	\kappa_{\eta\left(r\right)} =\frac{r}{r_\eta}.
\end{equation}
Here the parameter $\kappa_{\eta\left(r\right)}$ is numerically equal to the ratio of the number of absorbed photons of the wave in the laser-stimulated Compton process $\left(r\geq1,2,3...\right)$ to the characteristic quantum parameter $r_\eta$, which is determined by the experimental conditions and the laser installation.
\begin{equation} \label{eq3}
	r_\eta =\frac{\left(mc^2\right)^2\left(1+\eta^2\right)}{4\left(\hbar \omega\right)E_i\sin^2\left(\theta_i/2\right)}.
\end{equation}
Here $\theta_i$ is the angle between the momentum of the initial electron and the direction of wave propagation.  It can be seen from expression \eqref{eq3} that the value of the characteristic parameter $r_\eta$ is inversely proportional to the photon energy of the wave $\left(\hbar \omega \right)$ and the energy of the initial electrons $\left(E_i\right)$, and is also directly proportional to the intensity of the wave $\left(I\sim \eta^2 \left(\text{Wcm}^{-2}\right)\right)$.
It is important to note that the value of the parameter $\kappa_{\eta\left(r\right)}$ significantly affects the resonant frequency of the spontaneous gamma quantum \cite{33}.  So, for $\kappa_{\eta\left(r\right)}\ll1$ the resonant frequency is much less than the energy of the initial electron $\left(\hbar \omega'_{\eta\left(r\right)}\sim \kappa_{\eta\left(r\right)}E_i \ll E_i\right)$. If a parameter $\kappa_{\eta\left(r\right)}\sim1$, then the resonant frequency is of the same order with the energy of the initial electron $\left(\hbar \omega'_{\eta\left(r\right)}\sim E_i\right)$. If condition $\kappa_{\eta\left(r\right)}\gg1$ is met, then the resonant frequency will be close to the energy of the initial electron $\left(\hbar \omega'_{\eta\left(r\right)}\to E_i\right)$. At the same time, for a fixed value of the parameter $r_\eta$, the processes with a small number of absorbed photons of the wave $\left(r\sim1\right)$  have the largest resonant cross-section. As the number of absorbed photons of the wave increases $\left(r\gg1\right)$, the resonant cross-section decreases rapidly. It is important to emphasize that the production of high-energy spontaneous gamma quanta is possible only under the condition 
\begin{equation} \label{eq4}
	\kappa_{\eta\left(r\right)} \gg 1.
\end{equation}
The fulfillment of condition \eqref{eq4} is possible in two areas of variation of the quantum parameter $r_\eta$ \eqref{eq3}. Thus, for $r_\eta\gtrsim1$ condition \eqref{eq4} is fulfilled only for a large number of absorbed photons of the wave $r\gg1$, when the probability of these processes is significantly less than the probability of emission of low-energy spontaneous gamma quanta. This case was studied in detail in \cite{33}. On the other hand, when
\begin{equation} \label{eq5}
	r_\eta \ll 1
\end{equation}
the fulfillment of condition \eqref{eq4} already takes place for a small number of absorbed photons of the wave $\left(r\geq1,2,3...\right)$, when the probability of these processes will be maximum. This case was not considered in \cite{33}.

In this paper, we study the case of radiation of narrow high-energy gamma-ray beams under appropriate experimental conditions for a quantum parameter $r_\eta$ \eqref{eq5}.

We will use the relativistic system of units: $\hbar=c=1$.

\section{The Amplitude SB}

\begin{figure}[H]
	\centering
	\includegraphics[width=15cm]{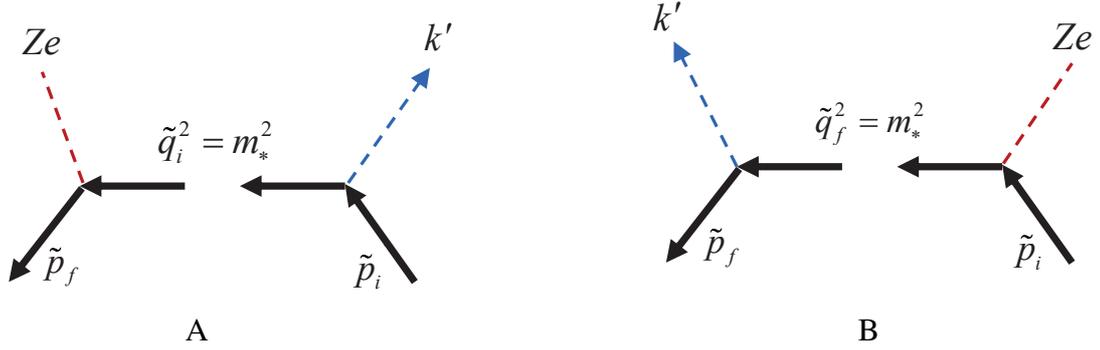} \\   A \qquad \qquad \qquad \qquad \qquad \qquad \qquad \qquad \qquad  \qquad  B
	\caption{Resonant spontaneous bremsstrahlung of an electron in the field of a nucleus and a plane electromagnetic wave.}
	\label{figure1}
\end{figure}

We choose the 4-potential of a plane monochromatic circularly polarized electromagnetic wave propagating along the axis $z$ in the following form:
\begin{equation} \label{eq6}
	A\left( \phi  \right)=\frac{F}{\omega }\left( e_x \cos \phi +\delta \cdot e_y \sin \phi  \right),\quad \phi =kx=\omega \left( t-z \right).
\end{equation}
Here $\delta =\pm 1$, $e_{x,y}=\left( 0,\mathbf{e}_{x,y} \right)$ and $k=\left( \omega ,\mathbf k \right)$ are 4-polarization vectors and the 4-momentum of the photon of the external field, and: $k^2=0,\ e^2_{x,y}=-1,\ e_{x,y}k=0$. This is a second-order process with respect to the fine structure constant and is described by two Feynman diagrams (see Figure \ref{figure1}). 
\unskip
The wave functions of the electron are determined by the Volkov functions \cite{53,54,55,56,57}, the intermediate states of the electron are given by the Green function in the field of a plane light wave \eqref{eq6} \cite{54,55}. The amplitude of such a process after simple calculations can be represented in the following form (see, for example, \cite{33}): 
\begin{equation} \label{eq7}
	S_{fi}=\sum_{l=-\infty }^{\infty } S_l,
\end{equation}
where the partial amplitude with the emission and absorption of $|l|$-photons of the wave has the following form:
\begin{equation} \label{eq8}
	S_l=-i\cdot \frac{8\pi^{5/2}\cdot Ze^3}{\sqrt{2\omega '\tilde E_i\tilde E_f}}\cdot \exp \left( i \varphi_{fi} \right)\cdot \left[ \bar u_f M_l u_i \right]\cdot \frac{\delta \left(q_0\right)}{\mathbf q^2}.
\end{equation}
Here it is indicated:
\begin{equation}\label{eq9}
	M_l=\sum \limits_{r=-\infty }^{\infty }   \left[  M_{r-l} \left( \tilde p_f, \tilde q_i \right) \cdot \left( \frac {\hat q_i+m} {\tilde q_i^2-m_*^2} \right) \cdot F_{-r} \left( \tilde q_i, \tilde p_i \right) + F_{-r} \left( \tilde p_f, \tilde q_f \right) \cdot \left( \frac{\hat q_f+m}{\tilde q_f^2-m_*^2} \right)\cdot M_{r-l} \left( \tilde q_f, \tilde p_i \right) \right].
\end{equation}
In Exps. \eqref{eq8}-\eqref{eq9} $\varphi_{fi}$ is the phase independent of the summation indices, $u_i$,$\bar u_f$ are Dirac bispinors, $\tilde p_i$ and $\tilde p_f$ are the 4-quasimomenta of the initial and final electrons, $\tilde q_i$ and $\tilde q_f$ are the 4-quasimomenta of intermediate electrons for channels A and B, $m_*$ is the effective mass of the electron in the plane wave field, $q$ is the 4-transmitted momentum:
\begin{equation}\label{eq10}
	\tilde q_i=\tilde p_i-k'+rk,\ \tilde q_f=\tilde p_f+k'-rk.
\end{equation}	
\begin{equation}\label{eq11}
	q=\tilde p_f-\tilde p_i+k'-lk.
\end{equation}
\begin{equation}\label{eq12}
	\tilde p_j=p_j+\eta^2 \frac{m^2}{2\left( kp_j \right)}k,\ \tilde q_j=q_j+\eta^2 \frac {m^2}{2\left( kq_j \right)}k,\quad j=i,f.
\end{equation}
\begin{equation}\label{eq13}
	\tilde p^2_{i,f}=m^2_*,\ m_*=m\sqrt{1+\eta^2}.
\end{equation}
Here $k'=\omega'\left( 1,\mathbf{n'} \right)$ is the 4-momentum of the spontaneous gamma-quantum, $p_{i,f}=\left( E_{i,f}, \mathbf{p}_{i,f} \right)$ is the 4-momentum of the initial and final electrons. Expressions with a cap in the relation \eqref{eq9} and further mean the scalar product of the corresponding 4-vector with the Dirac gamma matrices: $\gamma^\mu=\left( \gamma ^0,\boldsymbol {\gamma } \right)$, $\mu =0,1,2,3$. For example, $\hat{\tilde{q}}_i=\tilde{q}_{i\mu }\gamma^\mu=\tilde{q}_{i0}\gamma ^0-\mathbf{\tilde q_{i}} \boldsymbol {\gamma}$. Amplitudes $M_{r-l}$ and $F_{-r}$ (see Fig. \ref{figure1}) in the relation \eqref{eq9} have the form:
\begin{equation}\label{eq14}
	M_{r-l}\left( \tilde p_2, \tilde p_1 \right)=a^0\cdot L_{r-l}\left( \tilde p_2,\tilde p_1 \right)+b_{-}^0\cdot L_{r-l-1}+b_{+}^0\cdot L_{r-l+1},
\end{equation}
\begin{equation}\label{eq15}
	F_{-r}\left( \tilde p_2,\tilde p_1 \right)=\left( a\varepsilon^* \right)\cdot L_{-r}\left( \tilde p_2,\tilde p_1 \right)+ \left( b_{-}\varepsilon^* \right)\cdot L_{-r-1}+\left( b_+\varepsilon^* \right)\cdot L_{-r+1}.
\end{equation}
In these expressions $\varepsilon^*_\mu$ is the 4-polarization vector of the spontaneous gamma-quantum, and the matrices $a^\mu,b^\mu_\pm$ are determined by the relations
\begin{equation}\label{eq16}
	a^\mu=\tilde \gamma^\mu+\eta^2\frac{m^2}{2\left( k\tilde p_1 \right)\left( k \tilde p_2 \right)}k^\mu\hat k,
\end{equation}
\begin{equation}\label{eq17}
	b^\mu_\pm=\frac{1}{4}\eta m\cdot \left[ \frac{\hat{\varepsilon}_\pm \hat k \gamma^ \mu}{\left( k \tilde p_2 \right)}+\frac{\gamma^\mu \hat k \hat{\varepsilon }_\pm}{\left( k\tilde p_1 \right)} \right],\ \hat \varepsilon _\pm=\hat e_x\pm i\delta \cdot \hat e_y.
\end{equation}
Special functions $L_{r-l}$ and $\ L_{-r}$, and their arguments are given by expressions \cite{35}:
\begin{equation}\label{eq18}
	L_{r'}\left( \tilde p_2,\tilde p_1 \right)=\exp \left( -ir' \chi_{\tilde p_2 \tilde p_1} \right)\cdot J_{r'}\left( \gamma_{\tilde p_2 \tilde p_1} \right),
\end{equation}
\begin{equation}\label{eq19}
	tg{\chi_{\tilde p_2\tilde p_1}}=\delta \cdot \frac{\left( e_y Q_{\tilde p_2 \tilde p_1} \right)}{\left( e_x Q_{\tilde p_2 \tilde p_1} \right)},\quad Q_{\tilde p_2 \tilde p_1}=\frac{\tilde p_2}{\left( k \tilde p_2 \right)}-\frac{\tilde p_1}{\left( k \tilde p_1 \right)},
\end{equation}
\begin{equation}\label{eq20}
	\gamma_{\tilde p_2 \tilde p_1}=\eta m\sqrt{-Q^2_{\tilde p_2 \tilde p_1}}.
\end{equation}
Note that, $\left( k \tilde p_{1,2} \right)=\left( k p_{1,2} \right)$ and also in the case of amplitudes $M_{r-l}\left( \tilde p_f,\tilde q_i \right)$ and $M_{r-l}\left( \tilde q_f,\tilde p_i \right)$ in Exps. \eqref{eq14}, \eqref{eq16}-\eqref{eq17} it is necessary to put $\tilde p_1\to \tilde q_i,\ \tilde p_2 \to \tilde p_f$ and $\tilde p_1 \to \tilde p_i,\ \tilde p_2\to \tilde q_f$, and for the amplitudes $F_{-r}\left( \tilde q_i,\tilde p_i \right)$ and $F_{-r}\left( \tilde p_f,\tilde q_f \right)$ in the Exps. \eqref{eq15}-\eqref{eq17}: $\tilde p_1 \to \tilde p_i,\ \tilde p_2 \to \tilde q_i$ and $\tilde p_1 \to \tilde q_f,\ \tilde p_2 \to \tilde p_f.$ We should note the obtained amplitude of the SB process \eqref{eq7}-\eqref{eq20} is valid for arbitrary frequencies and intensities of a circularly polarized electromagnetic wave.

\section{Poles of the SB Amplitude in a Strong Field}

The resonant behavior of the amplitude \eqref{eq8}-\eqref{eq9} is due to the quasi-discrete structure of the system: an electron + a plane electromagnetic wave, as a result of which the 4-quasimomentum of the intermediate electron, due to the implementation of the laws of conservation of energy-momentum in the components of the process, lies on the mass shell \cite{31,32,33}. Because of this, for channels A and B, we get:
\begin{equation}\label{eq21}
	\tilde q^2_i=m^2_*, \qquad	\tilde q^2_f=m^2_*.
\end{equation}
We will study the most interesting case of ultrarelativistic electron energies, when the spontaneous gamma-quantum and the final electron fly out in a narrow cone along the momentum of the initial electron. In this case, the direction of propagation of the wave lies far from the given narrow cone of particles (if the direction of propagation of the wave lies inside the narrow cone of particles, then the resonances disappear \cite{31,32,33}).
\begin{equation}\label{eq22}
	E_{i,f}>>m,
\end{equation}
\begin{equation}\label{eq23}
	\theta'_{i,f}=\measuredangle \left( \mathbf{k'},\mathbf p_{i,f} \right)<<1,\quad \theta_{if}=\measuredangle \left( \mathbf p_i,\mathbf p_f \right)<<1,	
\end{equation}
\begin{equation}\label{eq24}
	\theta'=\measuredangle \left( \mathbf{k'},\mathbf k \right)\sim 1,\quad \theta_{i,f}=\measuredangle \left( \mathbf k,\mathbf p_{i,f} \right)\sim 1.	
\end{equation}

It should be noted that in strong laser fields, when the classical parameter $\eta \gtrsim 1$, instead of the mass and energy of the electron, it is necessary to use the effective mass and quasi-energy \cite{33}. Therefore, the condition \eqref{eq22} must be replaced with the following one:
\begin{equation}\label{eq25}
	\frac{\tilde E_{i,f}}{m_*}\approx \frac{E_{i,f}}{m\sqrt{1+\eta ^2}}\left[ 1+\left( \eta \frac{m}{2E_{i,f}} \right)^2\frac{1}{\sin^2\left( \theta _{i,f}/2\; \right)} \right]
	\sim \begin{cases}
		E_{i,f}/m>>1,\quad  & \text{if}\quad \eta <<1  \\
		E_{i,f}/{\left( \eta m \right)}>>1,\quad & \text{if}\quad \eta \gtrsim 1  \\
	\end{cases}.
\end{equation}
From the second condition of the relation \eqref{eq25}, we obtain a restriction on the maximum intensity of the laser wave:
\begin{equation}\label{eq26}
	\eta << {\eta}_{\max}=\frac{E_{i,f}}{m}>>1.
\end{equation}

In this paper, we will consider the energies of the initial electrons $E_i\lesssim 10^3\ \text{GeV}$ and also in a wide range of photon energies of an electromagnetic wave  $\left(1 \text {eV}\lesssim \omega \lesssim 10^4 \text {eV}\right)$. At the same time, we will consider the intensities of the electromagnetic wave significantly less than the critical intensities of the Schwinger  $\left(I\ll I_* \sim 10^{29} \text{Wcm}^{-2}\right)$.

Expressions defining the 4-quasimomenta of the intermediate electrons $\tilde q_i$ and $\tilde q_f$ \eqref{eq10}, as well as the transmitted 4-momenta $q$ \eqref{eq11}, for channels A and B (see Figure~\ref{figure1}) in resonance, it is convenient to write as:
\begin{equation}\label{eq27}
	\tilde p_i+rk=\tilde q_i+k',
\end{equation}
\begin{equation}\label{eq28}
	q=\tilde p_f-\tilde q_i+\left( r-l \right)k
\end{equation}
and
\begin{equation}\label{eq29}
	\tilde q_f+rk=\tilde p_f+k',
\end{equation}
\begin{equation}\label{eq30}
	q=\tilde q_f-\tilde p_i+\left( r-l \right)k.
\end{equation}
Since both the equalities $\tilde p^2_i=\tilde q^2_i=m^2_*$ and $\tilde p^2_f=\tilde q^2_f=m^2_*,\ k^2={k'}^2=0,$ then equations \eqref{eq27} and \eqref{eq29} are valid only for $r\ge 1$.

We determine the resonant frequency $\left( \omega'_{\eta i\left( r \right)} \right)$ of the spontaneous gamma-quantum for channel A. We take into account the relations \eqref{eq22}-\eqref{eq23} in the resonant condition \eqref{eq27}. After simple calculations \cite{33}, we get 
\begin{equation}\label{eq31}
	x'_{\eta i\left( r \right)}=\left[ 1+\frac{\left(1+{\delta'^2_{\eta i}} \right)}{\kappa_{\eta \left(r\right)}}\right]^{-1},\quad x'_{\eta i\left( r \right)}=\frac{\omega'_{\eta i\left( r \right)}}{E_i}.
\end{equation}
Here it is indicated:
\begin{equation}\label{eq32}
	\delta'_{\eta i}=\frac{E_i\theta '_i}{m_*}=\frac{E_i\theta '_i}{m\sqrt{1+\eta^2}}.
\end{equation}

Note that in the ultrarelativistic parameter $\delta'_{\eta i}$  \eqref{eq32} in the case of strong fields is determined by the effective mass of the electron $m_*$ \eqref{eq13} and the quasi-energy $\tilde E_i$ \eqref{eq12}. We put $\tilde E_i\approx E_i$ by virtue of condition \eqref{eq26}. It can be seen from the expression \eqref{eq31} that the resonant frequency of a spontaneous gamma quantum is determined by the outgoing angle (ultrarelativistic parameter $\delta'_{\eta i}$ ), as well as the quantum parameter $\kappa_{\eta \left(r\right)}$  \eqref{eq2}, \eqref{eq3}. Note that the resonant radiation spectrum \eqref{eq31} is essentially discrete, since each value of the number of absorbed laser photons corresponds to its own resonant frequency of the spontaneous gamma quantum: $r \to \omega'_{\eta i \left(r\right)}$ \eqref{eq31}). The resonance spectrum of spontaneous radiation in the region  $r_\eta\gtrsim 1$ $\left(0<\kappa_{\eta \left(r\right)}<\infty\right)$  was studied in detail in the article \cite{33}.

Here we study the case when the quantum characteristic parameter  $r_\eta$ satisfies condition \eqref{eq5}. Under these conditions, the parameter $\kappa_{\eta \left(r\right)}$ will satisfy condition \eqref{eq4}. As a result, for not very large outgoing angles $\left(\delta'^2_{\eta i}\ll\kappa_{\eta \left(r\right)}\right)$ for channel A the resonant frequency of the spontaneous gamma quantum \eqref{eq31} will be close to the energy of the initial electron
\begin{equation}\label{eq33}
	x'_{\eta i\left( r \right)} \approx 1-\frac{\left(1+\delta'^2_{\eta i}\right)}{\kappa_{\eta \left(r\right)}} \approx 1 \ \left(\delta'^2_{\eta i}\ll \kappa_{\eta \left(r\right)}\gg 1\right).
\end{equation}
Thus, in this case, with a small number of absorbed photons of the wave,  resonant spontaneous gamma quanta of maximum frequency will be emitted. The condition \eqref{eq5} can be rewritten for the energies of the initial electrons
\begin{equation}\label{eq34}
	E_i\gg E_*, \qquad E_*=\frac{m^2\left(1+\eta^2\right)}{4\omega \sin ^2 \left(\theta_i/2\right)}.
\end{equation}
Then, for the characteristic energy $E_*$  when the electron beam moves towards the electromagnetic wave $\left(\theta_i=\pi\right)$  and different frequencies and intensities of the wave, we obtain:
\begin{equation}\label{eq35}
	E_* = 6.5\cdot10^{10} \frac{\left(1+\eta^2\right)}{\omega\left(\text {eV}\right)}\text{eV}=
	\begin{cases}
		0.65 \text{GeV}, & \text{if} \ \ \omega=0.2 \text{keV}, I=0.746\cdot 10^{23} \text{Wcm}^{-2} \left(\eta=1\right)  \\
		6.5 \text {MeV}, & \text{if} \ \ \omega= \ 20 \text {keV}, I=0.746\cdot 10^{27} \text{Wcm}^{-2} \left(\eta=1\right)  \\
	\end{cases}.
\end{equation}
Consequently, for electron energies satisfying conditions \eqref{eq34}, \eqref{eq35} resonant spontaneous gamma quanta will be emitted with energies close to the energies of the initial electrons \eqref{eq33}.

We obtain the equation for the resonant frequency $\omega'_{\eta f \left(r\right)}$ of the spontaneous gamma-quantum in the case of channel B (see Figure~\ref{figure1}~B). Given the kinematics Eqs. \eqref{eq22}-\eqref{eq23}, from the expression \eqref{eq21} we get: 
\begin{equation}\label{eq36}
	\delta'^2_{\eta f} x'^3_{\eta f \left(r\right)}-2\delta'^2_{\eta f} x'^2_{\eta f \left(r\right)}+\left(1+\delta'^2_{\eta f} +\kappa_{\eta \left(r\right)} \right)x'_{\eta f \left(r\right)}-
	\kappa_{\eta \left(r\right)}=0, \  x'_{\eta f \left(r\right)}=\frac{\omega'_{\eta f \left(r\right)}}{E_i}.
\end{equation}
Here it is indicated
\begin{equation}\label{eq37}
	\delta'_{\eta f}=\frac{E_i \theta'_f}{m_*}=\frac{E_i \theta'_f}{m \sqrt{1+\eta^2}}.
\end{equation}
A detailed study of equation \eqref{eq36} for the case $r_\eta \gtrsim 1$ $\left(0<\kappa_{\eta \left(r\right)}<\infty\right)$  was carried out in  \cite{33}. Here we study the case when the quantum characteristic parameter $r_\eta$ satisfies condition \eqref{eq5}. Under these conditions, the parameter $\kappa_{\eta \left(r\right)}$ will satisfy condition \eqref{eq4}. As a result, for not very large outgoing angles $\left(\delta'^2_{\eta i}\ll\kappa_{\eta \left(r\right)}\right)$, from the equation \eqref{eq36}, it is easy to obtain the resonant frequency of a spontaneous gamma quantum for a channel B 
\begin{equation}\label{eq38}
	x'_{\eta f\left( r \right)} \approx 1-\frac{\left(1+\delta'^2_{\eta f}\right)}{\kappa_{\eta \left(r\right)}} \approx 1 \ \left(\delta'^2_{\eta f}\ll \kappa_{\eta \left(r\right)}\gg 1\right).
\end{equation}

\section{The Resonant Differential SB Cross Section }

It is important to emphasize that for the given initial parameters (the energy of the initial electron, the intensity and frequency of the wave, the angle between the momenta of the initial electron and the wave) for channel A, the energy of the spontaneous gamma quantum and the final electron is determined only by the outgoing angle of the spontaneous gamma quantum relative to the momentum of the initial electron (parameter $\delta'^2_{\eta i}$ , see Eq. \eqref{eq31}, \eqref{eq33}). At the same time, for channel B, the energy of the final particles is determined by the angle between the momenta of the spontaneous gamma-quantum and the final electron (the parameter $\delta'^2_{\eta f}$ , see Exps. \eqref{eq36}, \eqref{eq38}). Because of this, channels A and B are distinguishable and do not interfere with each other. In addition, within the same channel (A or B), resonances with a different number of absorbed photons of the wave (a different number $r$ ) have different frequencies and, therefore, also do not interfere. Taking this into account, it is possible to obtain a resonant differential cross section for each reaction channel separately at fixed values of the number of photons of the wave $l$  and $r$ . At the same time, in the sum by the   index in expression \eqref{eq9}, it is sufficient to leave a specific term with a fixed number of absorbed photons of the wave.

Using the expression for the process amplitude (see Exps. \eqref{eq7}-\eqref{eq9}), we obtain an expression for the resonant differential cross section in the case of unpolarized particles. After the standard calculations (see, for example, \cite{33}) for channel A, we get:
\begin{equation}\label{eq39}
	d\sigma_{i\left(l,r\right)}=d M_{i\left(r-l\right)}\frac{8 \pi ^2m^2E_f}{\left| \tilde q^2_i-m^2_* \right|^2}d K_{i\left(r\right)}.
\end{equation} 
Here $d M_{i\left(r-l\right)}$ is the differential cross-section of the scattering of an intermediate electron with a 4-momentum $\tilde q_i$ on the nucleus with the emission or absorption of $\left| r-l \right|$- photons of the wave \cite{50}:
\begin{equation}\label{eq40}
	d M_{i\left(r-l\right)}=4Z^2r^2_e\frac{m^2}{\mathbf q^4}J^2_{r-l}\left( \gamma_{\tilde p_f, \tilde q_i} \right)\delta \left[ \tilde E_f- \tilde q_{i0}+\left( r-l \right)\omega  \right]d^3\tilde p_f.
\end{equation}
\begin{equation}\label{eq41}
	\mathbf q=\mathbf{\tilde p}_f-\mathbf{\tilde q}_i+\left( r-l \right)\mathbf k,
\end{equation}
\begin{equation}\label{eq42}
	\gamma_{\tilde p_f, \tilde q_i}=\eta m\sqrt{-Q^2_{\tilde p_f, \tilde q_i}},\quad Q_{\tilde p_f, \tilde q_i}=\frac{\tilde p_f}{\left( k \tilde p_f \right)}-\frac{\tilde q_i}{\left( k \tilde q_i \right)}.
\end{equation}
Function $d K_{i\left(r\right)}$ determines the differential probability (per unit of time) of the laser-stimulated Compton effect with the absorption of $r$-photons of the wave \cite{44}:
\begin{equation}\label{eq43}
	d K_{i\left(r\right)}=\frac{\alpha }{4\pi \omega'E_i}K\left( u_{\eta i\left( r \right)},v_{\eta \left( r \right)} \right)d^3k',
\end{equation}
where
\begin{equation}\label{eq44}
	K\left(u_{\eta i\left( r \right)},v_{\eta \left( r \right)} \right)=-4J^2_r\left( \gamma_{\eta i\left( r \right)} \right)+\eta^2 \left(2+\frac{u^2_{\eta i\left( r \right)}}{1+u_{\eta i\left( r \right)}} \right) \left( J^2_{r+1}+J^2_{r-1}-2J^2_r \right),
\end{equation}
\begin{equation}\label{eq45}
	\gamma_{\eta i\left( r \right)}=2r\frac{\eta }{\sqrt{1+\eta ^2}}\sqrt{\frac{u_{\eta i\left(r\right)}}{v_{\eta \left(r\right)}}\left( 1-\frac{u_{\eta i\left( r \right)}}{v_{\eta \left( r \right)}} \right)},
\end{equation}
\begin{equation}\label{eq46}
	u_{\eta i\left(r\right)}=\frac{\left(kk'\right)}{\left(kq_i\right)}\approx \frac{x'_{\eta i\left( r \right)}}{\left(1-x'_{\eta i\left(r\right)} \right)}, \quad v_{\eta \left(r\right)}=2r\frac{\left( kp_i \right)}{m^2_*}=\kappa_{\eta \left(r\right)}.
\end{equation}
Here $\alpha $ is the fine structure constant. Note that, due to condition \eqref{eq26}, in the cross section \eqref{eq40}, it is possible to put $d^3\tilde{p}_f\approx d^3p_f\approx E^2_fdE_fd\Omega_f$ and easily carry out the integration with respect to the energy of the final electron. The elimination of the resonant infinity in channels A and B can be achieved by an imaginary addition to the effective mass of the intermediate electron \cite{33,56}. So, for channel A, we have:
\begin{equation}\label{eq47}
	m_*\to \mu_*=m_*+i\Gamma_{\eta i\left( r \right)},\quad \Gamma_{\eta i\left( r \right)}=\frac{\tilde q_{i0}}{2m_*}W\left( r_\eta \right).
\end{equation}
Here $W\left( r_\eta \right)$ is the total probability (per unit time) of the laser-stimulated Compton effect on an intermediate electron with the 4-momentum $\tilde q_i$ \cite{44}.
\begin{equation}\label{eq48}
	W\left( r_\eta \right)=\frac{\alpha m^2}{4\pi E_i} \mathrm K \left( r_\eta  \right),
\end{equation}
where
\begin{equation}\label{eq49}
	\mathrm K \left(r_\eta \right)=\sum_{s=1}^\infty \mathrm K_s \left(r_\eta \right),\quad \mathrm K_s \left(r_\eta \right)=\int\limits_0^{s/r_\eta} \frac{du}{\left(1+u\right)^2} K \left(u, \frac{s}{r_\eta} \right).
\end{equation}
Given the relations \eqref{eq48}, \eqref{eq49}, the resonant width \eqref{eq47} will take the form:
\begin{equation}\label{eq50}
	\Gamma_{\eta i\left( r \right)}=\alpha m\frac{\left( 1-x'_{\eta i\left( r \right)} \right)}{8\pi\sqrt{1+\eta^2}}\mathrm K \left( r_\eta \right).
\end{equation}
Here the function $K\left( u,s/r_\eta \right)$ is obtained from the expressions \eqref{eq44}-\eqref{eq46} by replacing: $u_{\eta i\left( r \right)}\to u,\ r\to s$. Taking into account the relations \eqref{eq47}-\eqref{eq50}, the resonant denominator can be represented as:      
\begin{equation}\label{eq51}
	\left| \tilde q^2_i-\mu^2_* \right|^2=m^4x'^2_{\eta i\left( r \right)}\left[ \left( 1+\eta^2 \right)^2 \left( \delta'^2_{\eta i}-\delta'^2_{\eta i\left( r \right)} \right)^2+\Upsilon^2_{\eta i\left( r \right)} \right].
\end{equation}
Here $\Upsilon_{\eta i\left( r \right)}$ is the angular resonant width.
\begin{equation}\label{eq52}
	\Upsilon_{\eta i\left( r \right)}=\frac{2\Gamma _{\eta i\left(r \right)}}{m_*x'_{\eta i\left( r \right)}}=\alpha \frac{\left(1-x'_{\eta i\left(r \right)}\right)}{4\pi x'_{\eta i\left( r \right)}} \mathrm K \left( r_\eta \right).
\end{equation}
In expression \eqref{eq51}, the $\delta'^2_{\eta i\left( r \right)}$ parameter is related to the resonant frequency by the ratio \eqref{eq31}, and the $\delta'^2_{\eta i}$ parameter  changes independently of the frequency of the spontaneous gamma-quantum. Note that the angular resonance width $\Upsilon _{\eta i\left( r \right)}$ \eqref{eq52} increases significantly with increasing wave intensity. Because of this, the angular resonance width $\Upsilon _{\eta i\left( r \right)}$  will be significantly greater than the corresponding radiation corrections \cite{37}. Taking this into account, we will carry out the corresponding integrations at the outgoing angles of the spontaneous gamma quantum (for channel A by parameter $\delta'^2_{\eta i}$  and for channel B by parameter $\delta'^2_{\eta f}$ ). Finally, the resonant differential cross section with simultaneous registration of the outgoing angle and frequency of the spontaneous gamma quantum for channels A and B will take the form (see detailed calculations in \cite{33}).
\begin{equation}\label{eq53}
	\frac{d\sigma_{\eta i\left(l,r\right)}}{dx'_{\eta i\left( r \right)}d\delta'^2_{\eta i}}=\left( Z^2 \alpha r^2_e \right)\frac{E^2_i}{m^2_*} \frac{4\pi^3 \left( 1-x'_{\eta i\left( r \right)} \right)} {g_{\eta i \left(r\right)}x'_{\eta i\left( r \right)}\left[ \left( 1+\eta^2 \right)^2 \left(\delta'^2_{\eta i}-\delta'^2_{\eta i\left( r \right)} \right)^2+\Upsilon^2_{\eta i\left( r \right)} \right]} K\left( u_{\eta i\left( r \right)}, \kappa_{\eta \left(r\right)}\right),
\end{equation}
\begin{equation}\label{eq54}
	\frac{d\sigma_{\eta f\left( l,r \right)}}{dx'_{\eta f\left( r \right)}d\delta'^2_{\eta f}}=\left( Z^2\alpha r^2_e \right)\frac{E^2_i}{m^2_*} \frac{4\pi^3 \left( 1-x'_{\eta f\left( r \right)} \right)^{-1}} {g_{\eta f \left(r\right)}x'_{\eta f\left( r \right)}\left[ \left( 1+\eta^2 \right)^2 \left(\delta'^2_{\eta f}-\delta'^2_{\eta f\left( r \right)} \right)^2+\Upsilon^2_{\eta f\left( r \right)} \right]} K\left( u_{\eta f\left( r \right)}, \kappa_{\eta \left(r\right)}\right).
\end{equation}
Here the relativistic-invariant parameter $u_{\eta f\left( r \right)}$ and the resonant width for channel B have the form:
\begin{equation}\label{eq55}
	u_{\eta f\left( r \right)}=\frac{\left( kk' \right)}{\left( kp_f \right)}\approx \frac{x'_{\eta f\left( r \right)}}{\left( 1-x'_{\eta f\left( r \right)} \right)},
\end{equation}
\begin{equation}\label{eq56}
	\Upsilon _{f\left( r \right)}=\frac{\alpha }{4\pi x'_{\eta f\left( r \right)}\left( 1-x'_{\eta f\left( r \right)} \right)}\mathrm K \left( r_\eta \right).
\end{equation}
The $K\left( u_{\eta f\left( r \right)}, \kappa_{\eta \left(r\right)}\right)$ function is obtained from the corresponding expression for channel A (see Eqs. \eqref{eq44}-\eqref{eq45}), if a replacement is made in the latter $u_{\eta i\left( r \right)}\to u_{\eta f\left( r \right)}$. The $g_{\eta i\left(r \right)},\ g_{\eta f\left(r \right)}$ functions  determine the square of the momentum transmitted to the nucleus, taking into account the relativistic corrections of the order $m^2_*/E^2_i$ for channels A and B. So, for channel A, we get:
\begin{equation}\label{eq57}
	g_{\eta i\left(r \right)}=g_{\eta i\left(r \right)}^{\left( 0 \right)}+\frac{1}{\left( 1+\eta^2 \right)}g_{\eta i\left(r \right)}^{\left( 1 \right)}+\frac{1}{\left(1+\eta^2\right)^2}g_{\eta i\left( r \right)}^{\left( 2 \right)},
\end{equation}
\begin{equation}\label{eq58}
		g_{\eta i\left(r \right)}^{\left( 0 \right)}=\left( 1-x'_{\eta i\left( r \right)} \right)+\frac{1}{\left( 1-x'_{\eta i\left( r \right)} \right)}+\frac{x'_{\eta i\left( r \right)}\delta'^4_{\eta i}}{6}\left[ \frac{1}{\left( 1-x'_{\eta i\left( r \right)} \right)^3}-1 \right]
		+\frac{\beta'_{\eta \left( r \right)}}{2\sin ^2\left( \theta_i/2 \right)}\left[ \beta '_{\eta \left( r \right)}-\frac{x'_{\eta i\left( r \right)}\left( 1+x'_{\eta i\left( r \right)} \right)}{\left( 1-x'_{\eta i\left( r \right)} \right)^2}\delta'^2_{\eta i} \right],
\end{equation}
\begin{equation}\label{eq59}
		g_{\eta i\left(r \right)}^{\left( 1 \right)}=\frac{x'_{\eta i\left( r \right)}}{\left( 1-x'_{\eta i\left( r \right)} \right)}\left[ \beta'_{\eta \left( r \right)}+\frac{x'_{\eta i\left( r \right)}}{\left( 1-x'_{\eta i\left( r \right)} \right)}\delta'^2_{\eta i} \right],\quad g_{\eta i\left( r \right)}^{\left( 2 \right)}=\frac{x'_{\eta i\left( r \right)}}{2}\left[ 1+\frac{x'_{\eta i\left( r \right)}\left( 2-x'_{\eta i\left( r \right)} \right)^2-1}{\left( 1-x'_{\eta i\left( r \right)} \right)^3} \right],
 \end{equation}	
\begin{equation}\label{eq60}
	\beta'_{\eta \left( r \right)}=\left( \frac{\eta^2}{1+\eta^2} \right)\frac{x'_{\eta i\left( r \right)}}{\left( 1-x'_{\eta i\left( r \right)} \right)}-\frac{r}{r_\eta}.
\end{equation}
Note that for channel B, the $g_{\eta f\left(r \right)}$ functions are obtained from $g_{\eta i\left(r \right)}$ functions \eqref{eq57}-\eqref{eq60} by replacing: $x'_{\eta i\left( r \right)}\to x'_{\eta f\left( r \right)},\ \delta'_{\eta i}\to \tilde \delta'_{\eta f}$. When the conditions are met
\begin{equation}\label{eq61}
	\left( \delta'^2_{\eta i}-\delta'^2_{\eta i\left( r \right)} \right)^2<<\frac{\Upsilon^2_{\eta i\left( r \right)}}{\left(1+\eta^2 \right)^2},\quad
	\left( \delta'^2_{\eta f}-\delta'^2_{\eta f\left( r \right)} \right)^2 <<\frac{\Upsilon^2_{\eta f\left( r \right)}}{\left( 1+\eta^2 \right)^2}
\end{equation}
we obtain the maximum resonant differential cross-sections for channels A and B:
\begin{equation}\label{eq62}
	R^{\max}_{\eta i\left( r \right)}=\frac{d\sigma^{\max}_{\eta i\left( r \right)}}{dx'_{\eta i\left( r \right)}d\delta'^2_{\eta i}}=\left(Z^2\alpha r^2_e \right)c_{\eta i}\Psi_{\eta i\left( r \right)},
\end{equation}
\begin{equation}\label{eq63}
	R^{\max}_{\eta f\left( r \right)}=\frac{d\sigma^{\max}_{\eta f\left( r \right)}}{dx'_{\eta f\left( r \right)}d\delta'^2_{\eta f}}=\left(Z^2\alpha r^2_e \right)c_{\eta i}\Psi_{\eta f\left( r \right)},
\end{equation}
Here, the $\Psi_{\eta i\left( r \right)}$ and $\Psi_{\eta f\left( r \right)}$ functions determine the spectral-angular distribution of the resonant SB cross-section for channels A and B:
\begin{equation}\label{eq64}
	\Psi_{\eta i\left( r \right)}=\frac{x'_{\eta i\left( r \right)}}{\left(1-x'_{\eta i\left( r \right)}\right)g_{\eta i\left( r \right)}}K\left( u_{\eta i\left( r \right)}, \kappa_{\eta \left( r \right)} \right),
\end{equation}
\begin{equation}\label{eq65}
	\Psi_{\eta f\left( r \right)}=\frac{x'_{\eta f\left( r \right)}\left(1-x'_{\eta f\left( r \right)}\right)}{g_{\eta f\left( r \right)}}K\left( u_{\eta f\left( r \right)}, \kappa_{\eta \left( r \right)} \right),
\end{equation}
and $c_{\eta i}$ - the coefficient, which is determined by the parameters of the laser installation
\begin{equation}\label{eq66}
	c_{\eta i}=\pi \left( \frac{8\pi^2 E_i}{\alpha m_* \mathrm K \left( r_\eta \right)} \right)^2\approx3.67\cdot 10^8 \frac{E^2_i}{m^2 \left(1+\eta ^2\right) \mathrm K^2 \left( r_\eta \right)}>>1. 
\end{equation}
In expressions \eqref{eq64} and \eqref{eq65}, the resonant frequencies are determined by the ratios \eqref{eq33} and \eqref{eq38}, respectively.

\section{Results}

\begin{figure}[H]
		\includegraphics[width=8cm]{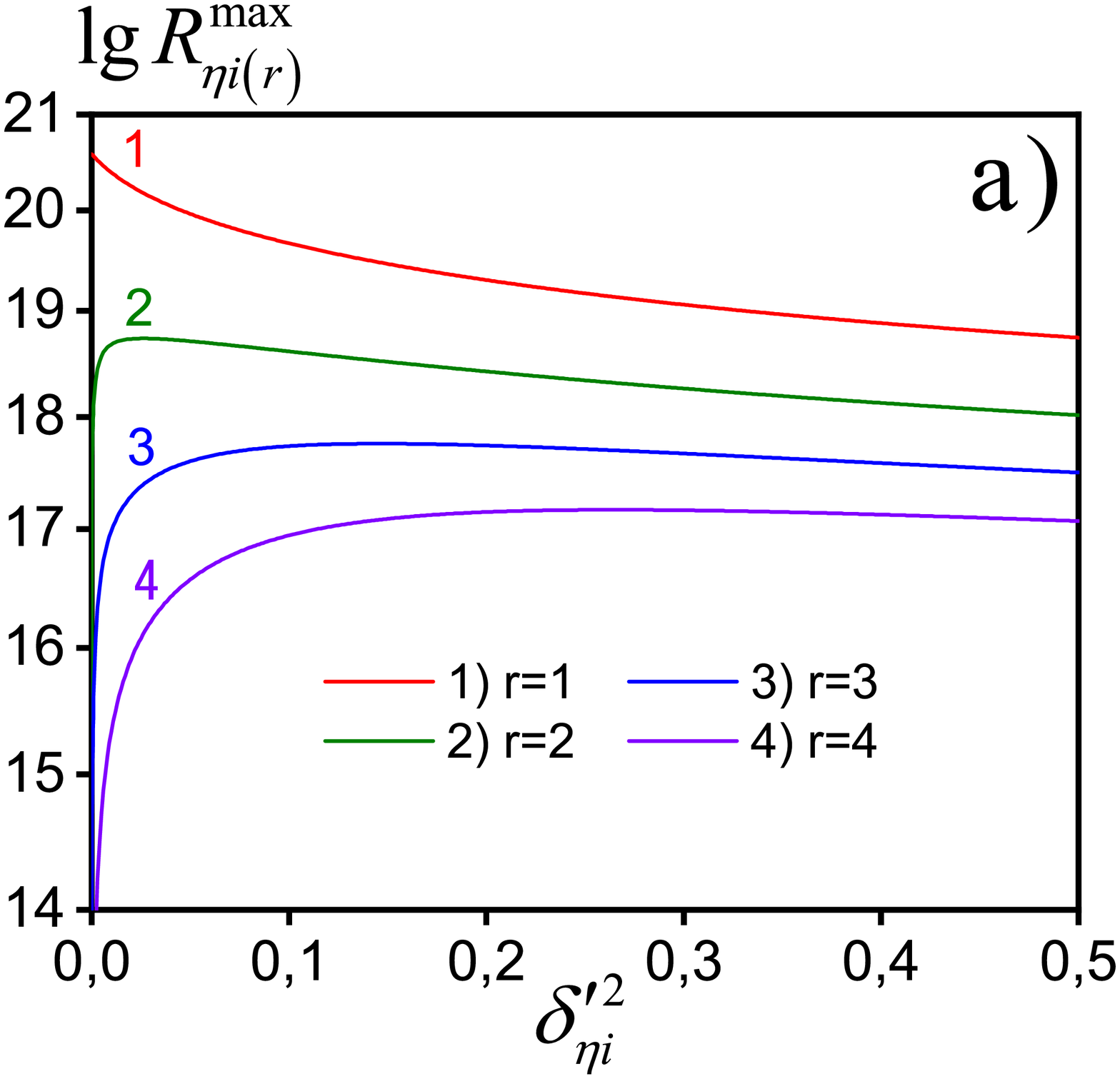}
		\includegraphics[width=8cm]{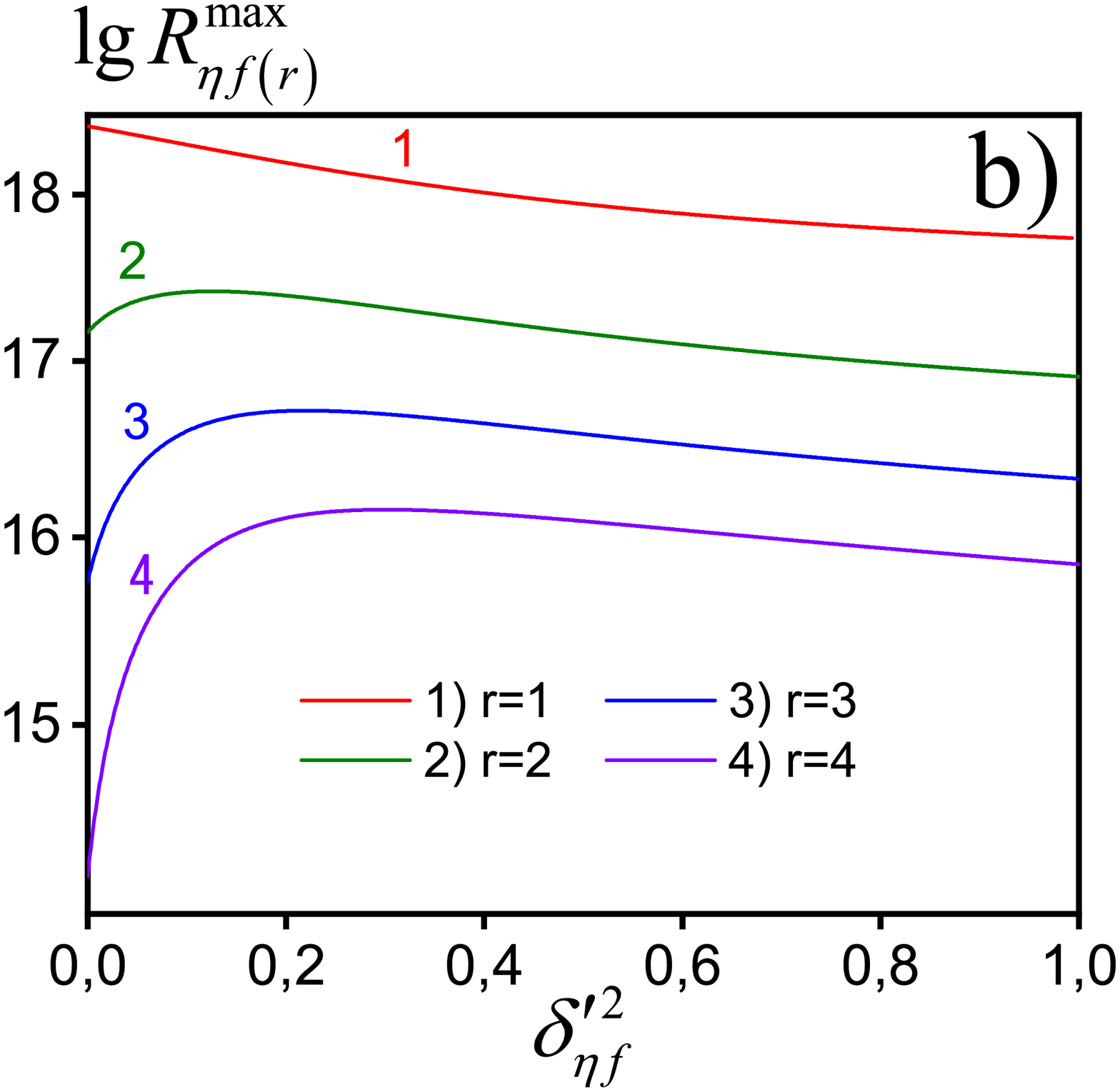}
	\caption{The maximum resonant differential cross-section (in units of $Z^2\alpha r^2_e$) as functions of the corresponding outgoing angles, plotted for the different values of absorbed wave photons.  (\textbf{a}) - $R^{\max}_{\eta i\left( r \right)}$ \eqref{eq62} for channel A. (\textbf{b}) - $R^{\max}_{\eta f\left( r \right)}$ \eqref{eq63} for channel B. The frequency and intensity of the electromagnetic wave are  $\omega=0.2\ \text{keV}$ and $I= 0.746\cdot 10^{23} \text{Wcm}^{-2}$. The energy of the initial electrons is $E_i=65\ \text{GeV}$.}
	\label{figure2}
\end{figure}

\begin{figure}[H]
		\includegraphics[width=8cm]{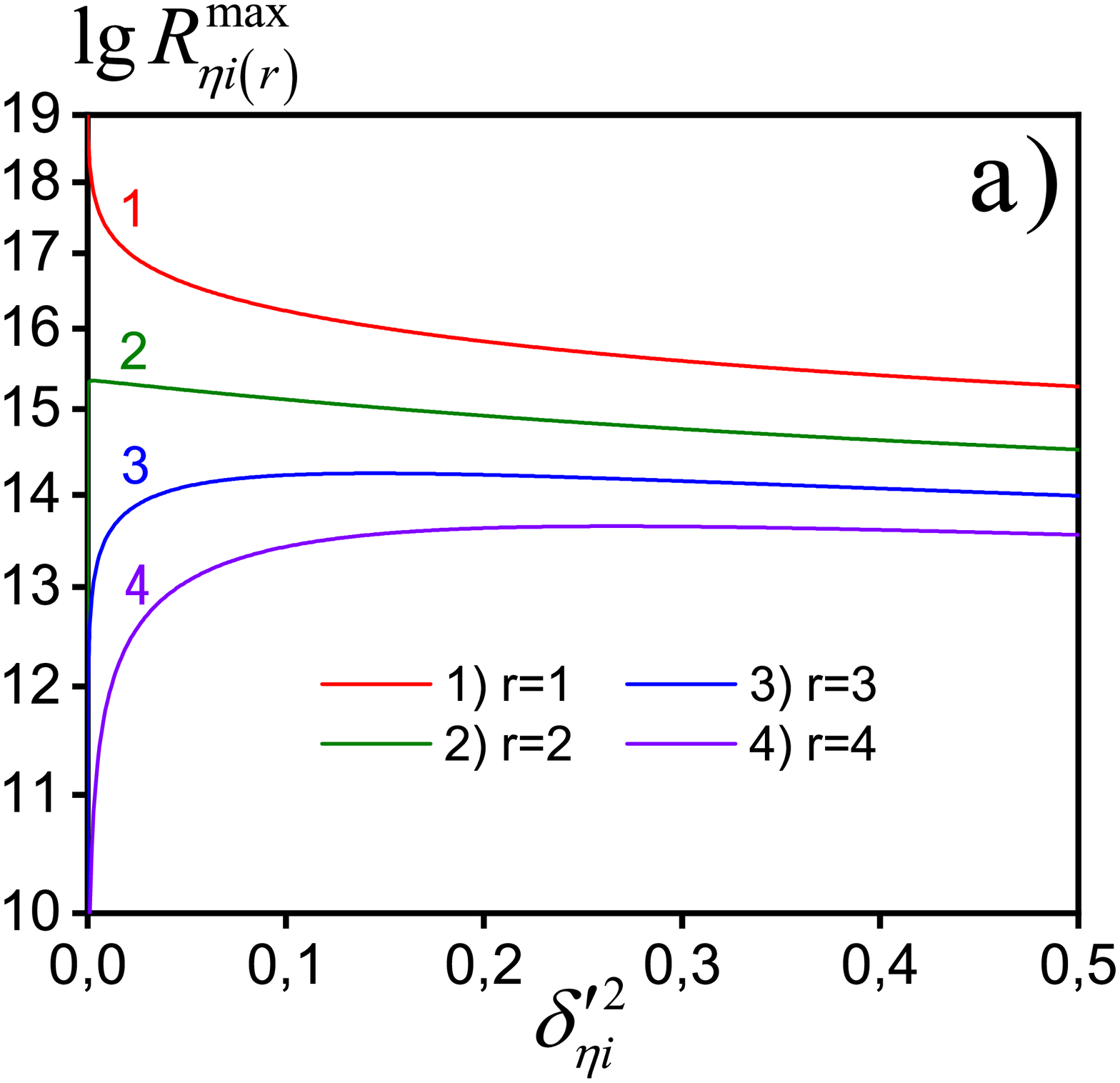}
		\includegraphics[width=8cm]{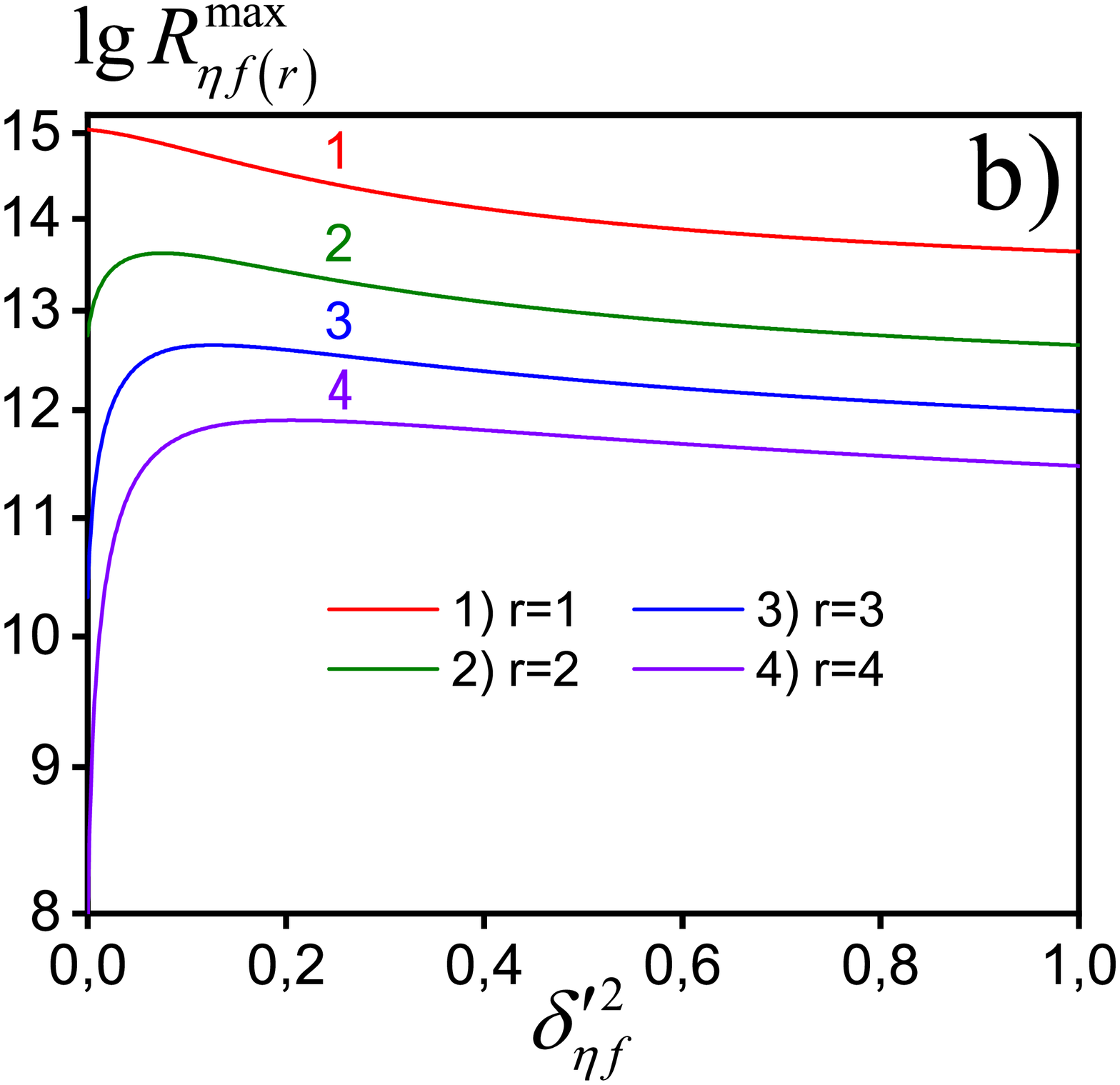}
	\caption{The maximum resonant differential cross-section (in units of $Z^2\alpha r^2_e$) as functions of the corresponding outgoing angles, plotted for the different values of absorbed wave photons.  (\textbf{a}) - $R^{\max}_{\eta i\left( r \right)}$ \eqref{eq62} for channel A. (\textbf{b}) - $R^{\max}_{\eta f\left( r \right)}$ \eqref{eq63} for channel B. The frequency and intensity of the electromagnetic wave are  $\omega=20\ \text{keV}$ and $I= 0.746\cdot 10^{27} \text{Wcm}^{-2}$. The energy of the initial electrons is $E_i=65\ \text{GeV}$.}
	\label{figure3}
\end{figure}

Figure~\ref{figure2},  Figure~\ref{figure3} - shows the distributions of the maximum resonant differential cross-section for channels A \eqref{eq62} and B \eqref{eq63} from the value of the square of the outgoing angle of the spontaneous gamma quantum with a different number of absorbed photons of the wave.

\unskip

\begin{table}[H] 
	\caption{The values of the resonant frequency and the square of the outgoing angle of the spontaneous gamma-quantum for the maximum value of the maximum resonant differential cross-section (see Figure~\ref{figure2}). The frequency and intensity of the electromagnetic wave are $\omega=0.2\ \text{keV}$ and $I= 0.746\cdot 10^{23} \text{Wcm}^{-2}$. The energy of the initial electrons is $E_i=65\ \text{GeV}$.\label{tab1}}
	\setlength{\extrarowheight}{2mm} 
	\begin{tabular}{|c|c|c|c|c|c|c|c|c|}
		\toprule
		$r$ & $\delta^2_{\eta i}$ & $R^{\max}_{\eta i\left( r \right)}$ & $\omega'_{\eta i\left( r \right)}$ & $E_{\eta f\left(r \right)}$ &$\delta^2 _{\eta f}$ & $R^{\max }_{\eta f\left( r \right)}$ & $\omega'_{\eta f\left( r \right)}$ & $E_{\eta f\left(r \right)}$\\
		& & $\left( Z^2\alpha r^2_e \right)$ & $\left(\text{GeV}\right)$ &  $\left(\text{GeV}\right)$ & & $\left( Z^2\alpha r^2_e \right)$ & $\left(\text{GeV}\right)$ & $ \left(\text{GeV}\right)$\\[2mm]
		\midrule
		1 & 0     & $1.147\times 10^{19}$ & 64.35644 &  0.64356  & 0     & $1.107\times 10^{15}$ & 64.35     &  0.65\\ \hline
		2 & 0.003 & $2.208\times 10^{15}$ & 64.67565 &  0.32435  & 0.075 & $4.159\times 10^{13}$ & 64.650625 &  0.349375\\ \hline
		3 & 0.145 & $1.749\times 10^{14}$ & 64.75287 &  0.24713  & 0.126 & $4.428\times 10^{12}$ & 64.756055 &  0.243945\\ \hline
		4 & 0.266 & $4.468\times 10^{13}$ & 64.79493 &  0.20507  & 0.205 & $8.008\times 10^{11}$ & 64.804155 &  0.195845\\ 
		\bottomrule
	\end{tabular}
	\setlength{\extrarowheight}{0mm} 
\end{table}
\unskip

\unskip

\begin{table}[H] 
	\caption{The values of the resonant frequency and the square of the outgoing angle of the spontaneous gamma-quantum for the maximum value of the maximum resonant differential cross-section (see Figure~\ref{figure3}). The frequency and intensity of the electromagnetic wave are $\omega=20\ \text{keV}$ and $I= 0.746\cdot 10^{27} \text{Wcm}^{-2}$. The energy of the initial electrons is $E_i=65\ \text{GeV}$.\label{tab2}}

	\setlength{\extrarowheight}{2mm} 
	\begin{tabular}{|c|c|c|c|c|c|c|c|c|}
		\toprule
		$r$ & $\delta^2_{\eta i}$ & $R^{\max}_{\eta i\left( r \right)}$ & $\omega'_{\eta i\left( r \right)}$ & $E_{\eta f\left(r \right)}$ &$\delta^2 _{\eta f}$ & $R^{\max }_{\eta f\left( r \right)}$ & $\omega'_{\eta f\left( r \right)}$ & $E_{\eta f\left(r \right)}$\\
		& & $\left( Z^2\alpha r^2_e \right)$ & $\left(\text{GeV}\right)$ &  $\left(\text{GeV}\right)$ & & $\left( Z^2\alpha r^2_e \right)$ & $\left(\text{GeV}\right)$ & $ \left(\text{GeV}\right)$\\[2mm]
		\midrule
		1 & 0     & $3.125\times 10^{20}$ & 64.99344 &  0.006565  & 0     & $1.172\times 10^{12}$ & 64.993435 & 0.006565\\ \hline
		2 & $2\times 10^{-5}$ & $6.343\times 10^{12}$ & 64.99675 &  0.00325  & 0.01 & $1.534\times 10^{10}$ & 64.996685 &  0.003315\\ \hline
		3 & 0.145 & $4.853\times 10^{11}$ & 64.99753 &  0.00247  & 0.028 & $2.806\times 10^{8}$ & 64.997725 &  0.002275\\ \hline
		4 & 0.266 & $1.239\times 10^{11}$ & 64.99792 &  0.00208  & 0.154 & $2.681\times 10^{7}$ & 64.998115 &  0.001885\\ 
		\bottomrule
	\end{tabular}
	\setlength{\extrarowheight}{0mm} 

\end{table}
\unskip

These figures differ in the frequencies $\left(\omega=0.2\ \text{keV};\ 20\ \text{keV}\right)$  and intensities of the electromagnetic wave $\left(I= 0.746\cdot 10^{23} \text{Wcm}^{-2}; \ 0.746\cdot 10^{27} \text{Wcm}^{-2}\right)$. Tables~\ref{tab1}~-~\ref{tab2} show the values of the resonant differential cross sections, energies and outgoing angles of the spontaneous gamma quantum, as well as the energies of the final electron in the maximum of the corresponding distributions, corresponding to Figures~\ref{figure2}~-~\ref{figure3}. It can be seen from Figure~\ref{figure2} –Figure~\ref{figure3} that at fixed frequency and intensity of the wave, the maximum value of the resonant section occurs at one absorbed photon of the wave $\left(r=1\right)$  for the zero outgoing angle $\left(\delta'^2_{\eta i}=0, \ \delta'^2_{\eta f}=0\right)$  and can be of the order of magnitude $10^{19}\div 10^{20} \ \left( Z^2\alpha r^2_e \right)$. With an increase in the number of absorbed photons of the wave, the magnitude of the maximum resonant cross-section decreases quite quickly, while still taking large values. At the same time, the position of the maxima in the corresponding distributions of resonant cross sections shifts towards large values of the outgoing angles of the spontaneous gamma quantum, and the energy of the spontaneous gamma quantum tends more and more to the energy of the initial electrons (see also Tables~\ref{tab1}~-~\ref{tab2}). Also note that channel B is suppressed relative to channel A.

\section{Conclusions}

The study of resonant spontaneous bremsstrahlung radiation during scattering of ultrarelativistic electrons by nuclei in the field of an X-ray wave with wide frequency and intensity intervals shows:
\begin{itemize}
	\item	Under conditions when the energies of the initial electrons significantly exceed the characteristic energy of the process ($E_i\gg E_*$ , see \eqref{eq34}), spontaneous gamma quanta will be emitted with energies close to the energy of the initial electrons (see \eqref{eq33}, \eqref{eq38}).
	\item	The emission peaks of spontaneous gamma quanta have different outgoing angles for different numbers of absorbed photons of the wave (see Figure~\ref{figure2} –Figure~\ref{figure3} and Table~\ref{tab1}-Table~\ref{tab2}).
	\item	The resonant differential cross-section with simultaneous registration of the outgoing angles and the energy of the spontaneous gamma quantum significantly exceeds the corresponding cross-section without an external electromagnetic field. This excess can be twenty orders of magnitude for one absorbed photon of the wave.
	\item	The resonant differential cross-section for channel A is dominant, since it exceeds the corresponding differential cross-section for channel B by more then two orders of magnitude.
\end{itemize}

The results obtained can be used to explain high-energy gamma-ray quanta near neutron stars and magnetars.


\vspace{6pt} 
 
\begin{thebibliography}{99}
	
	\bibitem{1} M. Khlopov, Astronomy Reports 59, 494 (2015).
	
	\bibitem{2} P. Goldreich and W.H. Julian, The Astrophysical Journal 157, 869 (1969).
	
	\bibitem{3} O. Adriani {\em et al.}, Nature 458, 607 (2009).
	
    \bibitem{4} D. Hooper, P. Blasi and P.D. Serpico, Journal of  Cosmology and Astroparticle Physics 2009, 025 (2009).
    
   	\bibitem{5} A. Dubov, V. V. Dubov and S. P. Roshchupkin, Universe 6, 143 (2020).
    
	\bibitem{6} N.R. Larin, S.P. Roshchupkin and V.V.Dubov, Universe 6, 141 (2020).
	
	\bibitem{7} D.V. Doroshenko, S.P. Roshchupkin and V.V.Dubov, Universe 6, 137 (2020).
	
	\bibitem{8} V.A. Yelatontsev, S.P. Roshchupkin and V.V.Dubov, Universe 6, 164 (2020).
	
	\bibitem{9} V.D. Serov, S.P. Roshchupkin and V.V.Dubov, Universe 6, 190 (2020).
	
	\bibitem{10} G.K. Sizykh, S.P. Roshchupkin and V.V.Dubov, Universe  7, 210 (2021).
	
	\bibitem{11} V. P. Oleinik, Zh. Eksp. Teor. Fiz. 52, 1049 (1967) [Sov. Phys. JETP 25, 697 (1967)].
	
	\bibitem{12} V. P. Oleinik, Zh. Eksp. Teor. Fiz. 53, 1997 (1967) [Sov. Phys. JETP 26, 1132 (1968)].
	
	\bibitem{13} F. Zhou and L. Rosenberg, Phys. Rev. {\bf A} 48, 505 (1993).
	
	\bibitem{14} A. V. Borisov, V. C. Zhukovskii, and P. A. Eminov, Sov. Phys. JETP 51, 267 (1980).
	
	\bibitem{15} M. Dondera and V. Florescu, Radiat. Phys. Chem. 75, 1380 (2006).
	
	\bibitem{16} A. Florescu and V. Florescu, Phys. Rev. {\bf A} 61, 033406 (2000).
	
	\bibitem{17} A. N. Zheltukhin, A. V. Flegel, M. V. Frolov, N. L. Manakov, and A. F. Starace, Phys. Rev. {\bf A} 89, 023407 (2014).
	
	\bibitem{18} A. V. Flegel, M. V. Frolov, N. L. Manakov, A. F. Starace, and A. N. Zheltukhin, Phys. Rev. {\bf A} 87, 013404 (2013).
	
	\bibitem{19} A. N. Zheltukhin, A. V. Flegel, M. V. Frolov, N. L. Manakov, and A. F. Starace, J. Phys. {\bf B} 48, 075202 (2015).

	\bibitem{20} A. Li, J. Wang, N. Ren, P. Wang, W. Zhu, X. Li, R. Hoehn, and S. Kais, J. Appl. Phys. 114, 124904 (2013).

	\bibitem{21} E. L$\mathrm{\ddot{o}}$tstedt, U. D. Jentschura, and C. H. Keitel, Phys. Rev. Lett. 98, 043002 (2007).

	\bibitem{22} S. Schnez, E. L$\mathrm{\ddot{o}}$tstedt, U. D. Jentschura, and C. H. Keitel, Phys. Rev. {\bf A} 75, 053412 (2007).
	
	\bibitem{23} A. A. Lebed' and S. P. Roshchupkin, Eur. Phys. J. {\bf D} 53, 113 (2009).
	
	\bibitem{24} A. Lebed' and S. Roshchupkin, Laser Phys. Lett. 6, 472 (2009).
	
	\bibitem{25} A. A. Lebed' and S. P. Roshchupkin, Phys. Rev. {\bf A} 81, 033413 (2010).
	
	\bibitem{26} S. P. Roshchupkin and O. B. Lysenko, Laser Phys. 9, 494 (1999).
	
	\bibitem{27} S. P. Roshchupkin and O. B. Lysenko, JETP 89, 647 (1999).
	
	\bibitem{28} A. A. Lebed', E. A. Padusenko, S. P. Roshchupkin, and V. V. Dubov, Phys. Rev. {\bf A} 94, 013424 (2016).
	
	\bibitem{29} A. A. Lebed', E. A. Padusenko, S. P. Roshchupkin, and V. V. Dubov, Phys. Rev. {\bf A} 97, 043404 (2018).
	
	\bibitem{30} S.P. Roshchupkin, N.R. Larin, and V.V. Dubov, Phys. Rev. {\bf D} 104, 116011 (2021).
	
	\bibitem{31} A. Dubov, V. V. Dubov and S. P. Roshchupkin, Laser Phys. Lett. 17, 045301 (2020).
		
	\bibitem{32} S. P. Roshchupkin, A. Dubov and V. V. Dubov, Laser Phys. Lett. 18, 045301 (2021).
	
	\bibitem{33} S.P. Roshchupkin, A.V. Dubov, V.V. Dubov and S.S. Starodub, New J. Phys. 24 (2022) 013020.	
	
	\bibitem{34} P. A. Krachkov, A. Di. Piazza, and A. I. Milstein, Phys. Lett. {\bf B} 797. 134814 (2019).
	
	\bibitem{35} S. P. Roshchupkin, V. A. Tsybul'nik, and A. N. Chmirev, Las. Phys. 10, 1256 (2000).
	
	\bibitem{36}  R. Kanya, Y. Morimoto, and K. Yamanouchi, Phys. Rev. Lett. 105, 123202 (2010).
	
	\bibitem{37}  A. Hartin, International Journal of Modern Physics {\bf A} 33, 1830011 (2018).
	
	\bibitem{38} S. P. Roshchupkin {\em et al.}, International Society for Optics and Photonics (SPIE, 2021) pp. 40 – 55. 
	
	\bibitem{39}  C. Bula, K. T. McDonald, E. J. Prebys, C. Bamber, S. Boege, T. Kotseroglou, A. C. Melissinos, D. D. Meyerhofer, W. Ragg, D. L. Burke $et \: al.$, Phys. Rev. Lett. 76, 3116 (1996).
	
	\bibitem{40} D. L. Burke, R. C. Field, G. Horton-Smith, J. E. Spencer, D. Walz, S. C. Berridge, W. M. Bugg, K. Shmakov, A. W. Weidemann, C. Bula $et \: al.$, Phys. Rev. Lett. 79, 1626 (1997).
	
	\bibitem{41} G. A. Mourou, T. Tajima, and S. V. Bulanov, Rev. Mod. Phys. 78, 309 (2006).
	
	\bibitem{42}   V. Bagnoud, B. Aurand, A. Blazevic, S. Borneis, C. Bruske, B. Ecker, U. Eisenbarth, J. Fils, A. Frank, E. Gaul $et \: al.$, Appl. Phys. {\bf B} 100, 137 (2010).
	
	\bibitem{43}  A. Di. Piazza, C. M$\mathrm{\ddot{u}}$ller, K. Z. Hatsagortsyan, and C. H. Keitel, Rev. Mod. Phys. 84, 1177 (2012).
	
	\bibitem{44} V. I. Ritus and A. I. Nikishov, Quantum electrodynamics phenomena in the intense field, in {\it Trudy FIAN}, edited by V. L. Ginzburg (Nauka, Moscow, 1979), Vol. 111.
	
	\bibitem{45} F. Ehlotzky, A. Jaro$\mathrm{\acute{n}}$, and J. Z. Kami$\mathrm{\acute{n}}$ski, Phys. Rep. 297, 63 (1998).
	
	\bibitem{46} F. Ehlotzky, K. Krajewska, and J. Z. Kami$\mathrm{\acute{n}}$ski, Rep. Prog. Phys. 72, 046401 (2009).
	
	\bibitem{47} S. P. Roshchupkin, Las. Phys. 6, 837 (1996).
	
	\bibitem{48} S. P. Roshchupkin, A. A. Lebed', and E. A. Padusenko, Las. Phys. 22, 1513 (2012).
	
	\bibitem{49} S. P. Roshchupkin, A. A. Lebed', E. A. Padusenko, and A. I. Voroshilo, Las. Phys. 22, 1113 (2012).
	
	\bibitem{50} M. V. Fedorov, {\it An Electron in a Strong Light Field} (Nauka, Moscow, 1991).
	
	\bibitem{51} S. P. Roshchupkin and A. I. Voroshilo, {\it Resonant and Coherent Effects of Quantum Electrodynamics in the Light Field} (Naukova Dumka, Kiev, 2008).
	
	\bibitem{52} S. P. Roshchupkin and A. A. Lebed', {\it Effects of Quantum Electrodynamics in the Strong Pulsed Laser Fields} (Naukova Dumka, Kiev, 2013).
	
	\bibitem{53} D. M. Volkov, Zeit. Phys. 94, 250 (1935).
	
	\bibitem{54} J. Schwinger, Phys. Rev. 82, 664 (1951).
	
	\bibitem{55} L. S. Brown and T. W. B. Kibblie, Phys. Rev. 133, A705 (1964).
	
	\bibitem{56} G. Breit and E. Wigner, Phys. Rev. 49, 519 (1936).
	
	\bibitem{57} V. B. Berestetskii, E. M. Lifshitz, and L. P. Pitaevskii, {\it Quantum Electrodynamics} (Nauka, Moscow, 1980).
	
	
\end{thebibliography}
\end{document}